# EFFECT OF MESOSCOPIC INHOMOGENEITIES ON THE CRITICAL CURRENT OF BULK MELT-TEXTURED YBCO.

L.S. Uspenskaya[1], I.G. Naumenko[1], G.A. Emelchenko[1], Yu.B. Boguslavskii[2], S.A. Zver'kov[1], E.B. Yakimov[3], D. Litzkendorf[4], W. Gawalek[4], A.D. Caplin[5]

[1] Institute of Solid State Physics Russian Academy of Science, Chernogolovka, Moscow distr, Russia, 142432

[2] General Physics Institute Russian Academy of Science, Moscow 117942, Russia

[3] Institute of Microelectronics Technology and High Purity Materials Russian Academy of Science, Chernogolovka, Moscow distr, Russia, 142432

[4] Institute of Physikalische Hochtechnologie, D-07702, Jena, Germany

[5] Blackett Laboratory, Imperial College, London SW7 2AZ, UK

**Abstract**

The downsizing *211*-inclusions and an increase of their density leads to rise in mean critical current value in Y-based melt textured material. Very often *211*-inclusion are spread in the material volume non-homogeneous, with typical scale *50 - 100* micrometer. Therefore it's difficult to find the real correlation between local critical current and the inclusions distribution. We performed a study of a local critical current using modified magneto-optic technique on a melt-textured *YBaCuO* ceramic, found the areas with constant current and studied the real structure of the material in the areas, inclusions distribution and their sizes, by scanning electron microscopy and X-ray microanalysis. The estimation of a pinning in these places, by taking into account the amount of inclusions and the length of their boundaries, and comparison with the value of local critical current reveals a strait correlation between the density of inclusions and the current but shows remarkable quantitative disagreement.

**Introduction.**

Bulk high temperature superconductor (HTSC) materials shaped as bars, rings, pellets and other are very promising for technological applications along with HTSC films and tapes [1]. Rotors and stators for electric motors and generators, elements for superconductor bearings and levitation systems, flywheels, quasi-permanent magnets for different applications could be produced from the bulk materials with high enough critical currents. The value of the current $J_c$ for such applications must be higher than $10^5 A/cm^2$ at $T = 77 K$. The highest $J_c$ is now achieved in the bulk melt-textured $YBa_2Cu_3O_x$ *(YBCO, Y123)* with small sized $Y_2BaCuO_5$ *(211)* inclusions [2]. The critical current is found to enhance with the diminishing of *211*-inclusion sizes. This size of precipitating *211* particles can be essentially diminished in the production process of bulk melt-textured *YBCO* by introducing of small amounts of $CeO_2$ and *Pt*. The amount and optimal size of non-superconducting inclusions required for obtaining a maximal $J_c$ in type 2 superconductors and, in particular, in melt-textured *YBCO* is the subject of wide speculation [3-23]. Perkins et al [24-25], Tirsa et al [26], based on the theory of collective vortex pinning, developed a method for generalization of experimental results for $J_c(B,T)$ for a considerable number of different *Re(123)* using only two parameters which govern the vortex pinning condition. But the physical essence of these parameters, their dependence on the real structure of *RE(123)* are not clear yet. This may be due to the fact that there is no exact correlation between $J_c(B,T)$ values and the real local *Re(123)* structure. In the largest number of the experimental works concerned with optimizing the composition and the mesoscale structure of *Re(123)* to reach the highest $J_c$, they do not take into account a local structural and composition inhomogeneities, with the characteristic size of *20-100 μm*, where the $J_c$ may vary several-fold. Naturally, the use of the mean values of $J_c$ and structure characteristics for a several millimeters samples blurs the physical picture of the correlation.

In this work we attempt to find the correlation of the local $J_c$ with the real *Y(123)* structure, with localization dictated by material inhomogeneities. We use a combination of macro and micro-structural methods, including electron microscopy and X-ray microanalysis, with modified magnetooptic method, where the magnetic fields alternately applied to a sample in different in-plane directions enable us to reveal areas, demonstrating different magnitudes of $J_c$ and pinning force (best, typical, not very bad superconducting areas).

**Experiment**

The melt-textured *Y(123)* samples were grown by technique, the details of which are described elsewhere [27], with a precursor composition *of $Y_{1.5}Ba_2Cu_3O_{7-x}$ + 1 wt% $CeO_2$*. The powder was uniaxially pressed. A modified MTG process using *MgO* seeds was applied to the blocks in a box furnace that allows heating and controlling the 6 sides. The process was performed with low temperature gradients ( $< 5\ K/cm$ ) and a cooling rate of *0.2 ÷ 1 K/h*. Finally the samples were oxygenated by a separate procedure. As-grown samples were 30 mm in diameter, 30mm high cylinders. Plate-like samples, 3x1x0.5 mm$^3$, were cut from a large domain on a top layer of the cylinder to draw the experiments.

The microstructure of samples was characterized using polarized optical and scanning electron JSM840 microscopes. The chemical composition was defined by a method of local X-ray spectral analysis using a crystal-diffraction X-ray spectrometer: X-ray microanalyzer of JXA-5- type (JEOL, Japan) equipped with a LINK AN 10/855 energy dispersion spectrometer. The accelerating voltage of *20 kV* was employed. The probe current was 0.*7 nÅ* on a metallic *Co*. The calibration was performed on the *CoKα* line. A mean chemical composition was defined by averaging the 5 μm diameter of the electron beam scanning over the area of *200x200 µm*. The composition of local regions in *Ce, Y, Ba, Cu, O* was defined with a *15 µm* diameter of the electron probe. In that case the accuracy of the elemental concentration determination did not exceed two relative percents.

The density of the whole as-grown *30 mm* cylinder defined by a hydrostatic weighing technique was *87 ± 0.1 %* of the theoretic density. The samples had $T_c = 92\ K$ with a transition width less than *2 K*, determined from the magnetic susceptibility temperature dependence.

A mean critical current in the sample was estimated from hysteresis loops. Mean intragrain current was determined by a standard magnetooptic technique [28] from magnetic field distribution profiles with allowance for real sizes of single-domain blocks. The scheme of the measurement is shown in Fig. 1a. The magnetoactive film placed on the flat polished sample surface, made it possible to observe and to measure the distribution of the magnetic field component, $H_n$, normal to that surface. In the experiment we magnetized the sample by the magnetic field applied in the same direction, perpendicular to the surface and estimated mean intragrain currents from the gradient of $H_n$ by means of the procedure described in [28].

Local critical current variations in the near surface layer of *20 - 100 µm* in depth were defined by a modified magnetooptic method. The measurement is given schematically in Fig. 2a. An external magnetic field $(H_{pl})$ was applied not perpendicular to the surface in the study, but parallel to it, i.e., in parallel with the superconductor sample surface. With such geometry of the experiment $H_n$ component results from variation of the local macroscopic current strength; in this sense the sample edges, or macroscopic surface roughness, show themselves only as regions of zero current. From variations of $H_n(x)$ along the direction *x* of the in-plane field $H_{pl}$, one can estimate changes in $J(x)$. In terms of Bean's model the current is determined by expression $J(x) = H_{pl}/4\pi\delta(x)$, where $\delta(x) = (1/H_{pl})\int H_n dx$. The spatial resolution of this method depends on the indicator film thickness and on the distance between the indicator and the sample surface. The sample surface was therefore carefully mechanically polished before the experiments. It is known that the indicator sensitivity to the normal component of the field (the rotation angle of polarized light under the action of the normal component of the field) decays as the longitudinal field is enhanced. At the same time, as the planar magnetic field is enhanced until it penetrates into the sample at a depth exceeding the characteristic size of inhomogeneties, the magnetooptic image contrast grows. This dictates an optimal planar field strength at which we perform our observations, enabling us to make up a map of variations of the critical current that occur in the near-surface *20 ÷ 100* micrometer layer, $J_c(x,y)$.

We made up the $J_c(x,y)$ map in most characteristic regions with minimal and maximal $J_c$ values and examined the composition and the structure of the material there. Before scanning electron microscopy (SEM), the samples surface was chemically etched in *3%$(NH_4)_2C_4O_6H_4$ + 3%$NH_4NO_3$* for several seconds to remove the oxidized near-surface layer [29].

**Experimental results.**

The outlook of the sample obtained in an optical microscope after a short chemical surface etching is illustrated in Fig. 3a. The optical observations reveal dark twinned areas of the *123* phase with a high density of inclusions (green *211* phase and others) and the light ones with a low density of inclusions. We failed to estimate the size and concentration of inclusions optically as they were too small. But the twinned structure was resolved fairly well in some regions, Fig. 3b. The twins were found to be oriented along the same directions within a few degrees in the whole of the sample plane. That means the crystallographic orientation was the same in the whole sample.

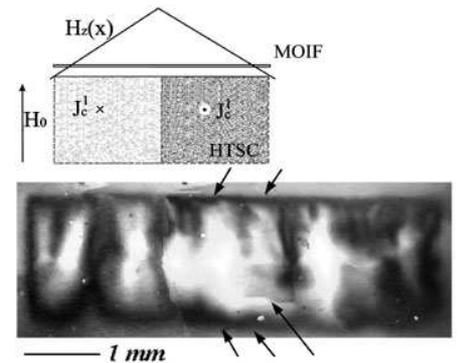

*Fig. 1. a) The scheme of standard magnetooptical observation of the magnetic flux distribution H(x) in a superconductor: the external magnetic field $H_0$ applied across observed surface induces a current flow in the bulk, MOIF – magnetic field sensitive magneto-optic indicator film is placed directly on the surface b) MO visualization of the magnetic flux trapped after normal 3 000 Oe magnetic field was applied and switched off (the brighter image, the higher magnetic field); macro and micro cracks conditioned main heterogeneity of the trapped magnetic flux, some heterogeneity are caused by structural variations (marked by arrows)*

The distribution of the normal to the sample surface magnetic field component $H_n$, obtained after cooling in the normal magnetic field $H_n = 1800\ Oe$ down to a temperature $T = 62\ K$ is illustrated in Fig. 1b. The flux is seen to partition to several groups despite the single block sample structure which is due to microcracks whose dotted traces are seen on the surface in an optical microscope. But the magnetic field gradient along the sample sides, proportional to the critical current, is varied even within microcrack-free regions, some of such regions are marked by arrows in Fig. 1b. We have determined the critical current, mean in a bulk sample, $J_c^{mean} \sim 1.1*10^4\ A/cm^2$ and mean in one-link sample areas, $J_c^{bl} \sim 3.4*10^4\ A/cm^2$ at $T = 77\ K$.

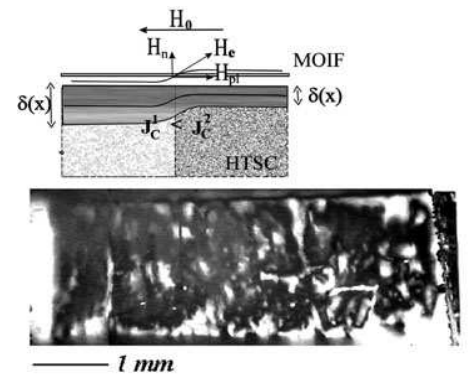

*Fig. 2. a) The scheme of modified magnetooptical observation of an inhomogeneous critical current distribution in a material: $H_0$ – an external magnetic field; $H_e$ - field, created by current flow in a superconductor, and it's components $H_n$, $H_{pl}$; $\delta(x)$ – the depth of the current flow; $J_c^i$ – local current; MOIF – magnetic field sensitive magneto-optic indicator film; b) MO visualization of the trapped magnetic flux induced by an in-plane field; scale of the magnetic field oscillations points at the scale of a local near surface critical current alteration.*

The distribution of $H_n$, obtained after cooling down to $T = 62\ K$ at the in-plane magnetic field $H_x = 1800\ Oe$ directed along the long sample side is illustrated in Fig. 2b. The estimations show that the field of such intensity is repelled from the sample depth of no more than $50 \div 100\ \mu m$. It is seen that not only the strength, but the field sign as well, vary along the surface with the characteristic scale $L \sim 20 \div 100$ μm. This suggests the current strength in a superconductor varies at a distance of the order of $L$. The analogous distribution of $H_n$ was also obtained for the field directed along the short sample side. We plotted then the $H_n(x)$, $H_n(y)$ profiles to estimate current strength variations along the surface and to distinguish the areas of $J_c^{best}$, $J_c^{aver}$, $J_c^{worse}$. We estimated the current strength variarions as $J_c^{best}:J_c^{aver}:J_c^{worse} \sim 1:0.8:0.5$ and $J_c^{best} \sim 10^5$ A/cm$^2$. It is very important to note, that crystallographic orientation of the compared areas is the same with the accuracy of *1* degree. Nevertheless the current in the areas differ twice.

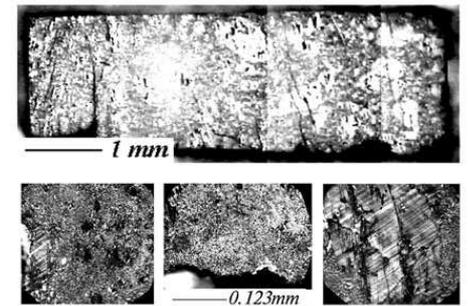

*Fig. 3. a) polarized light image of the sample surface: dark spots correspond to areas with high concentration of inclusions; b) optical images of some areas with larger magnification; the twins are seen as parallel lines, 211-inclusions are seen as tiny bit round shaped sports, the smallest of them could not be resolved by optic.*

As expected, the current was found much weaker in the areas with lower concentration of inclusions compared to those of higher one, as it is seen in an optical microscope. However the optical resolution is not enough to resolve the smallest inclusions. Therefore we used SEM to study the microstructure of the chosen areas. The SEM revealed the expected straight correlation between the concentration of inclusions in the superconducting *123* matrix and the critical current strength. In Fig. 4a is illustrated the structure of the best area, i.e., the area with the highest local current. We distinguished several such an areas in the sample. The light roundish spots in the figure are particles of the *211* phase, the gray background is the *123* matrix, the black sports are pores (they are in abundance) and "another" phase. The green *211* phase was identified by comparison the polarized optical image of largest green phase particles with their electron microscopic image. We failed to identify "another" phase with the available techniques because its size was too small. Minimal sizes of the phases are found as follows: *211* phase of *0.25* μ*m*, pores of *0.1* μ*m*, "another" phase of *0.25* μ*m*. The total volume concentration of the nonsuperconductive phases amounts to *20 ÷ 25 %*. The image of a typical area with good superconductivity is illustrated in Fig. 4b. In such areas the *211* inclusions sizes vary from *0.25* to *4* μ*m*. Small pores of minimally *0.15* μ*m* in size are observable but in small proportion. Along with inclusions and micropores there are linear defects in the form of fine bent lines. These can be either another unidentified phase or more likely stacking faults. The volume density of the *211* phase is even higher in such areas than in the areas with highest critical current; in individual areas the volume density of the *211* phase reaches 30% but inclusions are coarser on an average.

We found also the twinned areas with very small density of *211* phase concentration, Fig. 4c. They carry twice lower the critical current. Green *211* phase particles or pores are seldom encountered in such areas.

The dominant defects here are twin structures (diagonal lines in the figure) and compositional inhomogeneities (brightness modulation along the lines with a typical period of the order of *1 μm*). It should be mentioned that inhomogeneities of such type, the same as twins, are observable in all sample areas, including those with high concentration of inclusions. These defects could be caused by oxygen distribution variation in the *123* lattice, discussed elsewhere, e.g. [30].

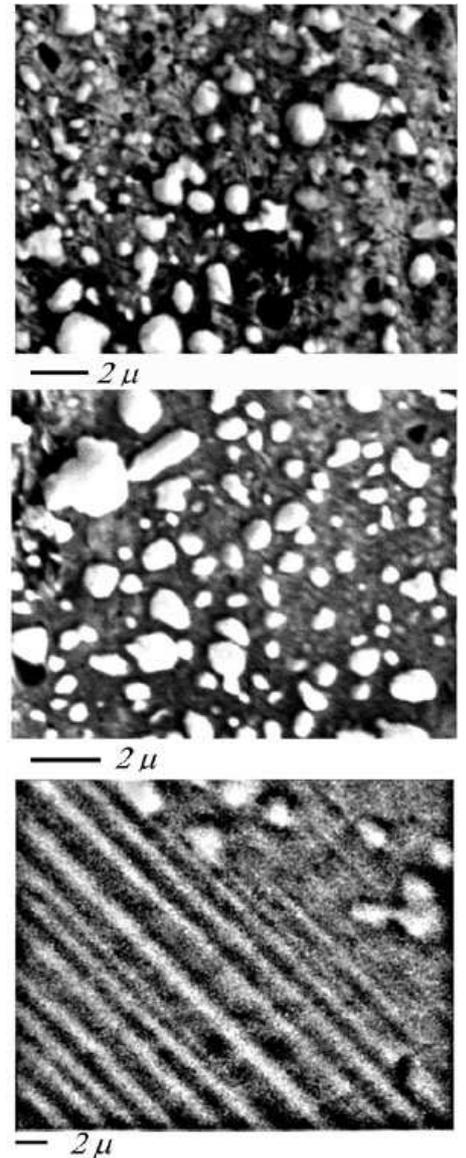

*Fig. 4. SEM image of the sample surface a) the area with the best superconducting properties; gray background – 123-matrix, white particles – 211-phase of 0.2 to 1 μm, black – pore space and "another phase" of down to 0.1 μm mainly round-shaped; b) the typical area with good superconducting properties; 211-phase of 0.25 to 4 μm, rare pores of down to 0.15 μm round-shaped and sub-micron long narrow stripe shaped inclusions or stacking faults; c) the typical twinned area of the same material with worse superconducting properties; twin are directed along the picture diagonal; white sports – rare 211-inclusions; variation of the contrast along the twins is probably due to the $BaCeO_3$ and the O inhomogeneous distribution*

So, we come to the common conclusion, there is a clear correlation between the type of the microstructure and the degree of magnetic flux pinning. In best areas the density of fine inclusions and pores is higher, compared to other areas. But the calculation of the vortex pinning on inclusions and pores visualized by us, with allowance for the density of inclusions and their sizes, [19, p.220] does not yield a current observed in our experiment. The disagreement between calculation and experiment was several orders of magnitude. That means that we miss some kind of very important pinning centers.

As mentioned above, we used chemical etching to prepare samples for SEM observation to remove near surface oxidized layer and get clear images. But chemical etching dissolves some small inclusions yielding additional pores; that was confirmed by comparison of SEM images of the same areas before chemical etching and after it. So, the real amount of pores in the material is less, than it is seen in the figures.

Table 1. Weigh concentration of *Ce, Y, Ba, Cu, O* in different areas of melt textured YBaCuO, determined by local X-ray spectral analysis.

|         | *Ce*      | *Y*    | *Ba*   | *Cu*      | *O*       |
|---------|-----------|--------|--------|-----------|-----------|
| Best    | *1±0.05*  | *21±1* | *38±1* | *23.4±0.1*| *16.3±0.1*|
| Typical | *0.65±0.05*| *19±5*| *43±5* | *25±3*    | *16.1±0.2*|
| Not bad | *0.5±0.1* | *18±5* | *39±5* | *25±3*    | *16±0.5*  |

We studied a chemical composition of the distinguished areas by X-ray microanalysis, Table 1. The samples were not chemically etched in this test. The measurements were performed in many sites of the areas in question and averaged, because of remarkable data scattering. The large spread in the obtained values is most likely due to the real local compositional inhomogeneities, than due to the accuracy of the equipment. We have found one more correlation along with the correlation between visible inclusions concentration and critical current density; the clear correlation between the local concentration of *Ce* and the local superconducting properties. The *Ce* concentration in the areas with the maximum current was twice higher than in the arrears with minimum current. We could not identify by SEM *Ce*-containing inclusions and determine their sizes exactly. We could only assume that *Ce* in the *123* matrix occurs, in part, as fine-disperse inclusions. This is validated by TEM observations made on another melt-textured material, $Y_1Ba_2Cu_3O_{7-x}$ + *2 wt% $ZrO_2$*, with the same value of the average critical current, as the *Ce*-containing *YBCO*. Large *Y123(Zr)* areas were found containing nonuniformly distributed fine inclusions of the order of *100 nm* in size and higher. The distances between these inclusions ranged from *100* to *200 nm*. The inclusions were identified as the *$BaZrO_3$* phase by a diffraction analysis. The sizes of *211* particles varied therewith from *300 nm* to tens of microns, in the same range as in *Y123(Ce)*. We believe that *$BaCeO_3$* is distributed in the *Y123(Ce)* material in the same manner as *$BaZrO_3$* in *Y123(Zr)*, and a highest *$BaCeO_3$* density is in the same areas where a highest concentration of submicron *211* inclusions is observed. These small-scale inclusions are more effective pinning centers than *211* particles. That is the reason why such high critical current as $J_c^{best} \sim 10^5$ *A/cm$^2$* is observed in some areas of the material.

**Discussion and conclusion.**

An electron-microscopic analysis of melt textured *YBCO* samples, containing *$Y_2BaCuO_5$, $BaCeO_3$, $BaZrO_3$* inclusions, performed in this work and the numerous reported data show that a minimal size of normal phase inclusions, being potential pinning centers, is about *100 nm; 211* inclusions observable in the melt-textured *Y123* are usually larger. Detailed analyses of theoretic predictions and, also, experimental data including those on fast-neutron bombardment of *YBCO* [19], an investigation of *YBCO(Ca)* superlattices [31], approximation of the empiric dependence $J_c \sim V_f / d$ for fine inclusions [19] give an evidence, that high density of critical current in *YBCO* ($J_c > 10^5$ *A/cm²* at *77 K*) could be achieved when the inclusions have the optimal sizes and distance between them: the size *d ~25nm*, the distance *L ~50nm* with volume fraction of such inclusions of about *6%*. Nowadays *211, Ag, BaZrO₃, BaCeO₃* and other material inclusions are used to create necessary pinning centers. We believe that *BaCeO₃* is practically feasible among other studied as it satisfy the following requirements: it is insoluble in *YBCO* melt, in the *Y123* lattice and does not suppress $T_c$; it is inert to the *YBCO* system and does not form new chemical compounds; it is not the nucleus for the *Y123* phase; it is enough refractory, obtainable as nanopowders with a mean particle size of the order of *25 nm*.

In conclusion, we have found some correlation between local $J_c$ and real microstructure of bulk melt-textured *Y123*. The density of pores of minimal size, *0.1 μm,* in best areas was observed to be comparable with that of fine *211* phase particles and it definitely correlates with the degree of the magnetic flux pinning suggesting, that small pores are real pinning centers in this material along with *211* inclusions. Our study shows that even 1% of fine spread *BaCeO₃* inclusions contributes effectively to the pinning in the local areas and enlarges remarkably the critical current.

We would like to thank A.Aronin, L.Zavelskaya, G.Panin for assistence in SEM study and fruitful discussions. The work was performed in the frame of Project #6458 and 6458A DMB50F, INTAS (project № 96 – 0251), Russian Foundation of Fundamental Reseach (project № 02-02-17062) and Russian Government contract №  40.012.1.1.11.46.